\documentclass[letter,twocolumn]{jpsj3}
\usepackage{graphicx}
\usepackage{color}
\usepackage{ulem}

\def\runauthor{K. Kudo {\it et al.}}

\title{Coexistence of Superconductivity and Charge Density Wave in SrPt$_2$As$_2$} 

\author{Kazutaka \textsc{Kudo}$^{1,2}$\thanks{E-mail address: kudo@science.okayama-u.ac.jp}, 
Yoshihiro \textsc{Nishikubo}$^{1,2}$, and Minoru \textsc{Nohara}$^{1,2}$}

\inst{$^1$Department of Physics, Faculty of Science, Okayama University, 3-1-1 Tsushima-naka, Kita-ku, Okayama  700-8530 \\ 
$^2$Transformative Research-Project on Iron Pnictides (TRIP), Japan Science and Technology Agency (JST), 5 Sanbancho, Chiyoda-ku, Tokyo 102-0075} 

\abst{
SrPt$_2$As$_2$ is a novel arsenide superconductor, which crystallizes in the CaBe$_2$Ge$_2$-type structure as a different polymorphic form of the ThCr$_2$Si$_2$-type structure. 
SrPt$_2$As$_2$ exhibits a charge-density-wave (CDW) ordering at about 470 K and enters into a superconducting state at $T_{\rm c} =$ 5.2 K. 
The coexistence of superconductivity and CDW refers to Peierls instability with a moderately strong electron-phonon interaction. 
Thus SrPt$_2$As$_2$ can be viewed as a nonmagnetic analog of iron-based superconductors, such as doped BaFe$_2$As$_2$, in which superconductivity emerges in close proximity to spin-density-wave ordering. 
}

\kword{non-iron-based pnictide, superconductivity, charge-density wave, SrPt$_2$As$_2$}

\begin{document}
\maketitle

Iron-based superconductors have been attracting considerable interest since Kamihara {\it et al.} reported superconductivity in LaFeAsO$_{1-x}$F$_x$ at $T_{\rm c} =$ 26 K\cite{rf:Kamihara}. 
A generic phase diagram of the iron-based superconductors shows that an antiferromagnetic (AFM) metallic and nonsuperconducting state is driven to a paramagnetic and superconducting state upon partial chemical substitution or by applying external hydrostatic pressure\cite{rf:Ishida,rf:Paglione}. 
BaFe$_2$As$_2$ is the compound most studied among the iron-based families so far and is widely thought to represent the generic features of the iron-based superconductors\cite{rf:Ishida,rf:Paglione}. 
BaFe$_2$As$_2$ exhibits AFM ordering below $T_{\rm N} =$ 140 K\cite{rf:Rotter1}. 
The partial chemical substitution of either Ba, Fe, or As ions with a different element suppresses the AFM ordering and induces superconductivity\cite{rf:Rotter2,rf:Chu,rf:Li,rf:Jiang}. 
The maximum superconducting transition temperature, $T_{\rm c} =$ 38 K for (Ba$_{1-x}$K$_x$)Fe$_2$As$_2$\cite{rf:Rotter2}, is obtained near the critical concentration where AFM ordering is completely suppressed.

It has been widely thought that AFM ordering in iron-based superconductors originates from the spin-density-wave (SDW) instability due to the nesting of two Fermi pockets, a hole pocket centered at the $\Gamma$ point and an electron pocket centered at the M point, which are connected by the nesting vector $Q = (\pi, \pi)$\cite{rf:Ding,rf:CLiu,rf:Brouet,rf:Terashima}. 
This characteristic electronic structure is believed to be a key ingredient of the high-temperature superconductivity due to AFM spin fluctuations in iron-based families\cite{rf:Ishida,rf:Paglione,rf:Mazin,rf:Kuroki}. 
Moreover, it has been theoretically proposed that the nesting determines the symmetry of Cooper pairs in the iron-based superconductors, in which a node locates either away from the Fermi surfaces (s$_\pm$-wave)\cite{rf:Mazin,rf:Kuroki} or directly at the Fermi surface (d-wave)\cite{rf:Kuroki}.

The charge-density-wave (CDW) instability is another consequence of Fermi-surface nesting when electron-phonon interaction remains important. 
CDW ordering, as well as SDW ordering, competes with or coexists with superconductivity\cite{rf:Gabovich,rf:Yomo,rf:Fang,rf:Singh,rf:Morosan,rf:Kiss}. 
In this letter, we demonstrate that superconductivity at $T_{\rm c} =$ 5.2 K coexists with CDW in platinum-based 122 arsenide SrPt$_2$As$_2$. 
The superconducting transition temperature of SrPt$_2$As$_2$ is considerably high compared with those of other non-iron-based 122 arsenides such as SrNi$_2$As$_2$ ($T_{\rm c} =$ 0.62 K)\cite{rf:Bauer} and BaNi$_2$As$_2$ ($T_{\rm c} =$ 0.7 K)\cite{rf:Ronning}. 
Our observation suggests that Fermi-surface nesting is a key ingredient of the relatively high superconducting transition temperature of SrPt$_2$As$_2$.

\begin{figure}[t]
\begin{center}
\includegraphics[width=0.95\linewidth]{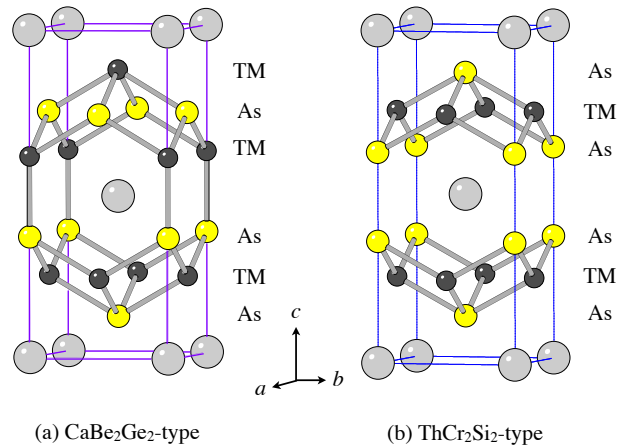}
\end{center}
\caption{
(Color online) Two polymorphic forms of 122: (a) CaBe$_2$Ge$_2$-type structure with the space group P4/nmm and (b) ThCr$_2$Si$_2$-type structure with the space group I4/mmm. 
The solid lines indicate the unit cell. 
The large circles and small circles represent Ba sites and Al sites of the BaAl$_4$-type structure with the space group I4/mmm. 
The Ba sites are occupied by Sr and the Al sites are occupied by either transition metal (TM) elements (represented by dark-gray circles) or arsenic (represented by yellow circles). 
}
\end{figure}
SrPt$_2$As$_2$ crystallizes in a tetragonal CaBe$_2$Ge$_2$-type structure with the space group P4/nmm (\#129)\cite{rf:Imre}, as shown in Fig. 1(a). 
The structure is different from another polymorphic form of 122, a ThCr$_2$Si$_2$-type structure with the space group I4/mmm (\#139), in which BaFe$_2$As$_2$ crystallizes\cite{rf:Rotter1,rf:Rotter2}, as shown in Fig. 1(b). 
Both structures are derived from the binary BaAl$_4$-type structure with the space group I4/mmm. 
In both forms, the barium sites are occupied by strontium atoms and the aluminum sites are occupied by either transition-metal (TM) atoms or arsenic atoms. 
The difference between the two forms is due to the different distributions of TM and arsenic over the aluminum sites. 
These structures can be schematically visualized by plane sequences along the $c$-axis: -As-TM-As-TM-As-TM- for the CaBe$_2$Ge$_2$-type structure and -As-TM-As-As-TM-As- for the ThCr$_2$Si$_2$-type structure. 
As can be seen from Fig. 1, there exists a three-dimensional Pt-As network in SrPt$_2$As$_2$ with the CaBe$_2$Ge$_2$-type structure, whereas there is a two-dimensional Fe-As network in BaFe$_2$As$_2$ with the ThCr$_2$Si$_2$-type structure.

Despite the three-dimensional Pt-As network, a band calculation indicates that the square lattice of Pt exhibits a Peierls instability in SrPt$_2$As$_2$\cite{rf:Imre}. 
In accordance with this prediction, SrPt$_2$As$_2$ exhibits a CDW transition at about 470 K. 
Imre {\it et al.} reported that a structural modulation develops with a modulation vector of $q = 0.62a^*$ in the -As-Pt-As- layers with PtAs$_4$ tetrahedra below about 470 K, whereas the -Pt-As-Pt- layers with Pt$_4$As tetrahedra remain intact. 
This modulation leads to a structural distortion from the high-temperature tetragonal phase with the CaBe$_2$Ge$_2$-type structure to the low-temperature CDW phase with the average structure of the orthorhombic space group Pmmn (\#59)\cite{rf:Imre}. 
In this phase, superconductivity emerges, as described in the following.

Polycrystalline samples of SrPt$_2$As$_2$ were synthesized by a solid-state reaction. 
PtAs$_2$ precursor was first synthesized by heating Pt powder and As grains at 700 $^\circ$C in an evacuated quartz tube. 
Then, stoichiometric amounts of Sr, PtAs$_2$, and Pt powders were mixed and ground. 
The resulting powder was placed in an alumina crucible and sealed into an evacuated quartz tube. 
The ampule was heated at 700 $^\circ$C for 10 h and then at 1100 $^\circ$C for 24 h. 
After furnace cooling, the sample was ground, pelletized, wrapped with Ta foil and heated at 700 $^\circ$C for 10 h in an evacuated quartz tube. 
The products were characterized by powder X-ray diffraction and confirmed to be a single phase of SrPt$_2$As$_2$. 
The lattice parameters were estimated to be $a = 4.46$  \AA, $b = 4.51$ \AA , and $c = 9.81$ \AA, which are consistent with the previous report\cite{rf:Imre}. 
This result indicates that the sample is indeed in the CDW phase at room temperature.

Magnetization $M$ was measured with a SQUID magnetometer (Magnetic Property Measurement System, Quantum Design) from 1.8 to 7 K under a magnetic field of 10 Oe. 
Electrical resistivity $\rho$ was measured by the standard DC four-terminal method in the temperature range between 2 and 300 K under magnetic fields up to 4 T using the Physical Property Measurement System (PPMS, Quantum Design). 
Specific heat $C$ was measured by the relaxation method in the temperature range between 2 and 7 K in zero field and in a magnetic field of 4 T using the PPMS.

Figure 2 shows the temperature dependence of the resistivity for a polycrystalline sample of SrPt$_2$As$_2$. 
The normal-state resistivity was of the order of 1 m$\Omega$cm at low temperatures. 
This relatively high value indicates that part of the Fermi surface is depleted by CDW formation. 
At high temperatures, the resistivity exhibited an `S'-shaped temperature dependence, which is characteristic of metal. 
We did not observe any anomaly in the normal-state resistivity below 300 K, indicating the absence of transition from the incommensurate CDW to a commensurate CDW. 
Thus, the incommensurate CDW with a modulation vector of $q = 0.62a^*$ persists down to low temperatures where superconductivity emerges. 
\begin{figure}[t]
\begin{center}
\includegraphics[width=1\linewidth]{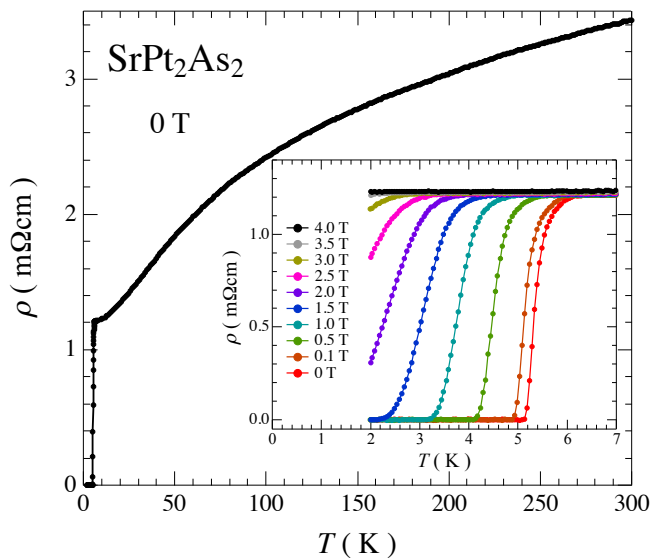}
\end{center}
\caption{
(Color online) Temperature dependence of electrical resistivity $\rho$ for SrPt$_2$As$_2$ in zero field. 
The inset shows temperature dependence of $\rho$ in magnetic fields up to 4 T. 
}
\end{figure}
\begin{figure}[t]
\begin{center}
\includegraphics[width=1\linewidth]{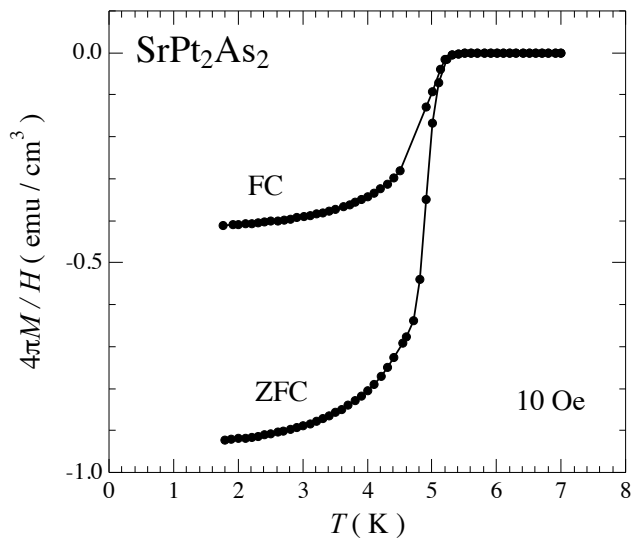}
\end{center}
\caption{
Temperature dependence of magnetization divided by applied field, $M/H$, of SrPt$_2$As$_2$ at 10 Oe under zero-field-cooling (ZFC) and field-cooling (FC) conditions. 
}
\end{figure}

The inset of Fig. 2 shows the resistive superconducting transition in detail. 
The 10$-$90 \% transition width was about 0.4 K, and the onset temperature determined from the 10 \% rule was 5.7 K. 
Zero resistivity was observed at 5.2 K. 
The temperature-dependent magnetization data for this sample are shown in Fig. 3, which exhibited a diamagnetic behavior below 5.2 K. 
The shielding and flux exclusion signals correspond to 92 and 42 \% of perfect diamagnetism, respectively. 
These data support the emergence of bulk superconductivity at $T_{\rm c} =$ 5.2 K in SrPt$_2$As$_2$.

As shown in  the inset of Fig. 2, $T_{\rm c}$ gradually decreased with increasing external magnetic field. 
The temperature dependence of the upper critical field  $H_{\rm c2}$ was determined from the midpoint of the resistive transition, as shown in Fig. 4. 
Interestingly, $H_{\rm c2}$ increased almost linearly with decreasing temperature down to the lowest temperature measured, and deviated from the curve expected from the Werthamer-Helfand-Hohenberg (WHH) theory\cite{rf:WHH}. 
Similar deviation of $H_{\rm c2}$ was reported for some multiband superconductors, such as YNi$_2$B$_2$C\cite{rf:Shulga} as well as doped-BaFe$_2$As$_2$\cite{rf:Ni}, but it is unclear at present whether SrPt$_2$As$_2$ is a multiband superconductor or not. 
Thus the WHH theory gives an estimate of the lower limit of the upper critical field at 0 K, $H_{\rm c2}(0) =$ 2.5 T, from the slope of $H_{\rm c2}$ at $T_{\rm c}$, $-dH_{\rm c2}/dT|_{T = T_{\rm c}} = 0.68$ T/K. 
The Ginzburg-Landau coherence length, $\xi_{\rm 0}$, was estimated to be 115 \AA \ from $\xi_{\rm 0} = [\Phi_{\rm 0}/2\pi H_{\rm c2}(0)]^{1/2}$, where $\Phi_{\rm 0}$ is the magnetic flux quantum. 
\begin{figure}[t]
\begin{center}
\includegraphics[width=1\linewidth]{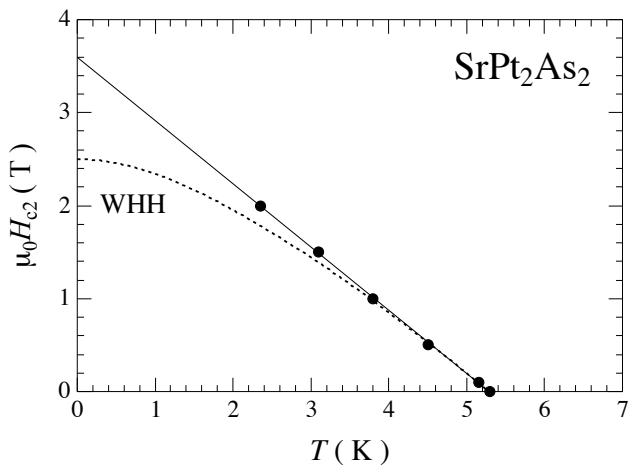}
\end{center}
\caption{
Temperature dependence of upper critical field $H_{\rm c2}$ deduced from resistivity measurements. 
The solid straight line yields a slope of $-dH_{\rm c2}/dT|_{T = T_{\rm c}} = 0.68$ T/K. 
The broken line represents a curve based on the Werthamer-Helfand-Hohenberg (WHH) theory\cite{rf:WHH}. 
}
\end{figure}

Further support of the bulk superconductivity was obtained from the specific heat $C(T)$, where a clear jump at the superconducting transition was observed, as shown in Fig. 5. 
In order to accurately determine bulk $T_{\rm c}$ in zero magnetic field, an ideal jump at $T_{\rm c}$ was assumed to satisfy the entropy conservation at the transition. 
This yielded estimates of $T_{\rm c} =$ 5.0 K and $\Delta C/T_{\rm c} =$ 16.2 mJ/molK$^2$. 
The superconductivity was completely suppressed in a magnetic field of 4 T. 
A standard analysis on the normal-state specific heat yielded the $T$-linear specific heat coefficient $\gamma =$ 9.72 mJ/molK$^2$ and Debye temperature $\Theta_{\rm D} =$ 211 K. 
Using these values, we estimated $\Delta C/\gamma T_{\rm c}$ to be 1.67, which is larger than the value expected from the BCS weak-coupling limit ($\Delta C/\gamma T_{\rm c} =$ 1.43).  
\begin{figure}[t]
\begin{center}
\includegraphics[width=1\linewidth]{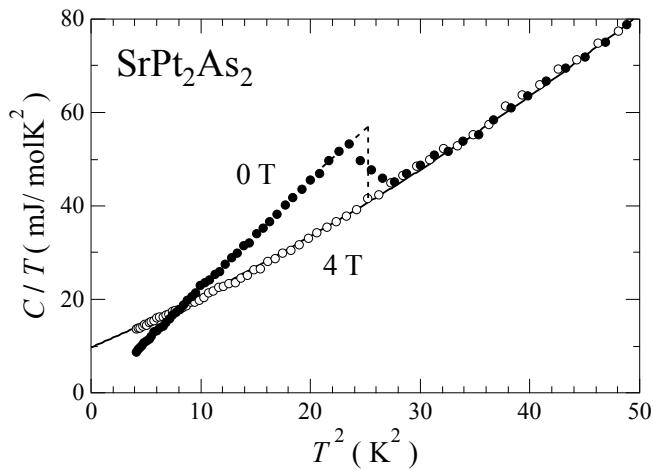}
\end{center}
\caption{
Specific heat divided by temperature, $C/T$, as a function of $T^2$ in zero field and a magnetic field of 4 T. 
The solid line represents a fit by $C/T = \gamma + \beta T^2 + \delta T^4$, where $\gamma$ is the coefficient of the electronic specific heat, while $\beta$ and $\delta$ are those of phonon contributions. 
The broken line represents an ideal jump at $T_{\rm c}$, assuming entropy conservation at the transition. 
}
\end{figure}

The $\gamma$ value of SrPt$_2$As$_2$ is comparable to those reported for other non-iron-based 122 superconductors\cite{rf:Bauer,rf:Ronning,rf:Ronning2,rf:Hirai,rf:Hirai2}. 
For the Ni-based arsenides, which are isoelectronic to SrPt$_2$As$_2$, $\gamma =$ 8.7 and 10.8 mJ/molK$^2$ for SrNi$_2$As$_2$\cite{rf:Bauer} and BaNi$_2$As$_2$\cite{rf:Ronning}, respectively. 
However, $T_{\rm c}$ is almost one order of magnitude different between SrPt$_2$As$_2$ and the Ni-based 122 system: $T_{\rm c} =$ 5.2 K for SrPt$_2$As$_2$, while $T_{\rm c} =$ 0.62  and 0.7 K for SrNi$_2$As$_2$\cite{rf:Bauer} and BaNi$_2$As$_2$\cite{rf:Ronning}, respectively. 
The difference in Debye temperatures, $\Theta_{\rm D} =$ 211 and 244 K for SrPt$_2$As$_2$ and SrNi$_2$As$_2$\cite{rf:Bauer}, respectively, is unable to account for the difference in $T_{\rm c}$ between the two classes of materials. 
We suggest that the relatively high $T_{\rm c}$ is most likely due to the enhanced electron-phonon interaction in SrPt$_2$As$_2$, as expected from the enhanced specific-heat jump in SrPt$_2$As$_2$, $\Delta C/\gamma T_{\rm c} =$ 1.67. 
The `S'-shaped temperature-dependent resistivity and its saturating behavior at high temperatures are consistent with the moderate electron-phonon coupling in SrPt$_2$As$_2$.

The present compound, SrPt$_2$As$_2$, and the iron-based superconductors share a common electronic structure with Fermi-surface nesting, which leads to Peierls instability. 
SrPt$_2$As$_2$ shows superconductivity in close proximity to CDW ordering due to electron-phonon coupling, while the iron-based families show superconductivity in close proximity to SDW/AFM ordering due to the electron correlations\cite{rf:Ishida,rf:Paglione,rf:Mazin,rf:Kuroki}. 
Because of the extremely wide diversity of materials in transition-metal pnictides, we may have a chance to develop a compound that possesses the characteristics of both platinum-based and iron-based pnictides. 
Such compounds will provide us an opportunity to realize novel superconductors with higher transition temperatures, since the high-$T_{\rm c}$ copper oxide superconductors show superconductivity closely related to both SDW/AFM and CDW/charge-ordered phases\cite{rf:Orenstein,rf:Kivelson,rf:Dagotto}.

In summary, SrPt$_2$As$_2$ is a novel pnictide superconductor undergoing both a charge-density-wave transition at 470 K and a superconducting transition at $T_{\rm c} =$ 5.2 K. 
The superconducting transition temperature of SrPt$_2$As$_2$ is considerably high among non-iron-based pnictide superconductors. 
The present superconductor, SrPt$_2$As$_2$, is analogous to iron-based arsenide superconductors in that the superconductivity emerges in close proximity to the Peierls instability.
We suggest that Fermi-surface nesting is a key ingredient for realizing high transition temperature.

\section*{Acknowledgement}
We would like to thank Dr. T. Nishizaki and Dr. N. Yoneyama for fruitful discussions. 
A part of this work was performed at the Advanced Science Research Center, Okayama University.

\end{document}